\documentclass[submission,copyright,creativecommons]{eptcs}
\usepackage{graphicx}
\usepackage{algorithm}
\usepackage[noend]{algpseudocode}
\usepackage{tikz}
\usepackage{caption}
\usepackage{subcaption}

\usepackage{amsmath,amsfonts}
\DeclareMathOperator{\atantwo}{atan2}

\usetikzlibrary{decorations.pathmorphing,intersections}
\usetikzlibrary{positioning,fit,calc}
\providecommand{\event}{SOS 2007} 

\usepackage{iftex}

\ifpdf
  \usepackage{underscore}         
  \usepackage[T1]{fontenc}        
\else
  \usepackage{breakurl}           
\fi

\author{Sergiy Bogomolov
\institute{Newcastle University, \\ Newcastle upon Tyne, United Kingdom}
\email{sergiy.bogomolov@ncl.ac.uk}
\and
Taylor T. Johnson
\institute{Vanderbilt University, \\ Nashville, USA}
\email{taylor.johnson@vanderbilt.edu}
\and
Diego Manzanas Lopez
\institute{Vanderbilt University, \\ Nashville, USA}
\email{diego.manzanas.lopez@vanderbilt.edu}
\and
Patrick Musau
\institute{Vanderbilt University, \\ Nashville, USA}
\email{patrick.musau@vanderbilt.edu}
\and
Paulius Stankaitis
\institute{Newcastle University, \\ Newcastle upon Tyne, United Kingdom}
\email{paulius.stankaitis@ncl.ac.uk}
}

\title{Online Reachability Analysis and Space Convexification for Autonomous Racing}
\def\titlerunning{Online Reachability Analysis and Space Convexification for Autonomous Racing}
\def\authorrunning{Bogomolov et al.}
\begin{document}
\maketitle

\begin{abstract}
This paper presents an optimisation-based approach for an obstacle avoidance problem within an autonomous vehicle racing context.  Our control regime leverages online reachability analysis and sensor data to compute the maximal safe traversable region that an agent can traverse within the environment. The idea is to first compute a non-convex safe region, which then can be convexified via a novel coupled separating hyperplane algorithm. This derived safe area is then used to formulate a nonlinear model-predictive control problem that seeks to find an optimal and safe driving trajectory. We evaluate the proposed approach through a series of diverse experiments and assess the runtime requirements of our proposed approach through an analysis of the effects of a set of varying optimisation objectives for generating these coupled hyperplanes.
\end{abstract}

\section{Introduction} 
\label{sec:mpcintroduction}

Over the last several years, autonomous racing has actively been pursued as a strategy to explore edge-case scenarios in autonomous driving \cite{Hartman}. Racing scenarios present unique challenges with respect to navigating high speeds and multi-agent interactions. In these contexts, vehicles must be able to operate at the edge of their operating envelopes in close proximity to static and dynamic obstacles. Several competitions have emerged over the last couple of years, such as the Indy Autonomous Challenge (IAC) \cite{Hartman}, and the F1TENTH International Autonomous racing competition \cite{okelly2020}. Although numerous racing strategies have been proposed over the last several years, head-to-head racing at high speeds remains a challenge. Unlike the time trials that are frequently used as qualification rounds in these competitions \cite{F1102019}, head-to-head racing requires designing a regime to be able to predict the future trajectories that reflect the intentions of the other opponents and drive through the track as quickly as possible. 

Within the autonomous racing space, one of the most popular frameworks for tackling the racing problem has been formulating and solving an optimisation problem that balances obstacle avoidance and travelling at high velocities \cite{Schoels2019,Katrakazas2015}. Specifically, the model-predictive control framework (MPC), which finds optimal control commands based on a model of the underlying system, while satisfying a set of constraints is the most widely used approach \cite{Jung2021}. Although MPC approaches have enjoyed success in these settings \cite{Schoels2019}, one of the main limitations exhibited by many approaches is a lack of robust online risk assessment in often dynamic and uncertain environments, particularly around vehicle-to-vehicle interactions. While a lot of progress has been made in this area, collisions still occur due to misplaced estimations of the set of all possible trajectories that the vehicle could pursue \cite{Katrakazas2015}. Furthermore, as Katrakazas et al. note ``exhaustively calculating and predicting the trajectories of other traffic participants at each epoch incurs a huge computational cost''.  Currently, many existing approaches treat the vehicle as an isolated entity, and the behavioural models of other participants within the environment have not yet been widely incorporated into the MPC regime \cite{Katrakazas2015}. 

One of the ways that this challenge has been addressed has been through the use of reachability analysis approaches \cite{Althoff2014}. The idea is to compute the set of states that the other racing agents could occupy in the future, for a fixed time horizon, and plan trajectories for the ego vehicle that avoids this unsafe set \cite{LeungReach2020,Liu2019,Lorenzetti2018}. This unsafe set allows for modelling the inherent uncertainty in the behaviour of other agents and for the synthesis of safe racing trajectories \cite{Althoff2014}. There are two main challenges that arise in these contexts. The first is that over long time horizons, reachability approaches will result in overly conservative behaviours as the set of avoidable states grows. The second is that reachability approaches are typically computationally challenging endeavours, thus leveraging them online is quite challenging. In light of these challenges, the following paper presents a model-predictive control framework leveraging real-time reachability for a 1/10 scale autonomous vehicle test-bed in a multi-agent racing setting modelled after the F1TENTH International Autonomous Racing Competition.

Finally, obtaining a solution to the MPC problem generally entails solving a convex optimisation problem, which guarantees convergence to a globally optimum solution. However, due to the presence of  static and dynamic obstacles, the optimal control problem of obstacle avoidance is inherently a non-convex problem. Therefore, to solve this problem efficiently many approaches leverage state-space convexification. In the past, several state-space convexification approaches have been proposed, including region partitioning \cite{doi:10.1080/00423114.2017.1399209}, computing separating hyperplanes \cite{Liniger2014,8022960}, and constructing approximations using stored data points \cite{rosolia2019learning} (further discussed in Section \ref{sec:mpcrelated}). In our framework, we propose  a novel optimisation-based approach for convexifying non-convex state spaces by computing coupled separating hyperplanes. The coupling of separating hyperplanes makes it possible to compute optimal safe and convex regions. However, it comes at the cost of increased computation time. Therefore, in this paper, we investigate the feasibility (e.g., timing constraints) of computing coupled separating hyperplanes in a real-time autonomous racing scenario.

In summary, the contributions of this paper are:
(1) we introduce a novel closed-loop model-predictive obstacle avoidance controller that integrates online reachability analysis and an optimisation-based state-space convexification approach, (2) we evaluate this approach across a diverse set of simulation experiments using the F1TENTH simulation platform. These experiments include varying the number of dynamic agents, the number of static and dynamic obstacles, and the racing environment. 
(3) We present a timing analysis of the state-space convexification approach. (4) Finally, we evaluate our approach against the well-known model-predictive contouring control approach, which has shown great success in obstacle avoidance tasks.

\section{Related Work}
\label{sec:mpcrelated}

Researchers have approached the obstacle avoidance problem from two major perspectives. The first strategy involved formulating and solving an optimisation problem. The second regime has typically involved a hierarchical decomposition of path planning and reference tracking. A variety of algorithms such as artificial potential fields \cite{Zhiyang}, genetic algorithms \cite{2008Yang}, rapidly-exploring random trees (RRT) \cite{Karaman}, fuzzy logic algorithms \cite{Mohammad2013}, elastic band theory \cite{Dong2021}, and rolling window methods \cite{Zhang2017} have demonstrated success in numerous arenas. A key limitation of many path planning approaches is that they are incapable of respecting kinodynamic constraints, such as bounds on the acceleration, and often the trajectories must be passed to a low-level controller that utilises a higher fidelity dynamics model and respects control constraints \cite{Schoels2019}. Furthermore, in highly dynamic and uncertain environments, planners must be able to replan sufficiently fast to react appropriately to split-second environmental threats \cite{Karaman}. However, most planners typically do not replan sufficiently rapidly to ensure split-second reactivity to threats \cite{LeungReach2020}.

As mentioned previously, MPC approaches have demonstrated great success in generating optimal trajectories that respect kinodynamic constraints and recently researchers have combined these approaches with reachability analysis to generate provably collision-free paths \cite{Althoff2014,Bajcsy2019Provably,LeungReach2020,Frazzoli2016,LeungReach2020}. Within this regime, \cite{Althoff2014,Frazzoli2016,Liu2019} utilise forward reachability methods in order to eliminate areas of the state space that would result in collisions. While these methods are extremely effective, these approaches must be implemented carefully in order to ensure that the resulting trajectories do not result in overly conservative behaviours \cite{LeungReach2020}. The alternative to these approaches is backward reachability approaches \cite{Bajcsy2019Provably,LeungReach2020} which utilise a target set representing a set of undesirable states, in order to design controllers that can guarantee dynamic and static obstacle avoidance with minimal intervention. However, these approaches are computationally demanding and typically the safety-ensuring control constraints, derived from these methods, are computed and cached offline before being incorporated into an MPC problem \cite{LeungReach2020}.

Beyond reachability methods, over the last several years, several space convexification approaches for the obstacle avoidance problem have been proposed. In \cite{7489011} a feasible convex set for model-predictive control is obtained by computing two parallel time-varying hyperplanes on racetrack borders. However, the resulting hyperplanes do not consider static obstacles or dynamic agents. The works of Mercy et al.~\cite{8022960,7810517} and  Scholte et al.~\cite{4480899}  utilise the concept of separating hyperplanes to compute hyperplanes which separate autonomous systems from convex obstacles. The paper  \cite{9525177} combines the model-predictive control and dynamic agent reachability analysis, and uses IRIS (Iterative Regional Inflation by Semi-definite programming) \cite{Deits2015} for a state-space convexification. A similar approach has been introduced in \cite{arxiv.1709.00627} for motion planning. Finally, in  \cite{doi:10.1080/00423114.2017.1399209} two (polar and convex) different types of convexification methods based on region partitioning for obstacle avoidance were proposed. Their convex partitioning regime utilises a convex partitioning algorithm \cite{doi:10.1142/S0218195902000803} to compute the minimum number of convex regions that are needed to capture non-convex obstacles, whereas the polar partitioning approach derives a safe set by using a minimum number of triangles.

The state-space convexification approaches described above have two main limitations for the racing scenario: they generally aim to compute the largest convex region in the non-convex space (e.g., not necessarily in the travelling direction of the ego vehicle) or are not able to handle non-convex obstacles. In our approach, we also express the problem of computing separating hyperplanes as an optimisation problem, but we are interested in computing a correct set of separating hyperplanes that provide the largest safe convex region in the direction of the ego vehicle. Furthermore, our proposed approach is able to handle non-convex obstacles.


\section{Preliminaries}
\label{sec:mpcpreliminaries}

\subsection{Model-Predictive Control}

Let us suppose we have the following (\ref{model:mpcsystem}) discrete-time system where ${x} \in {X} \subseteq \mathbb{R}^{n}$, ${u} \in {U} \subseteq\mathbb{R}^{m}$ and $t \in \mathbb{N}$.

\begin{equation}
    {x_{t+1}} = f({x_t, u_t})
    \label{model:mpcsystem}
\end{equation}

The MPC problem can then be expressed as a finite horizon optimisation problem (\ref{model:mpcsystem}) where a cost function $J$ is being minimised over a finite time horizon $N$ subject to constraints (2.1 - 2.4). 

\begin{align}
&J_{t \rightarrow t+N}({x}_t)=\min_{{u_0,\dots, u_{N-1}}} p({x}_{t+N})+\sum_{k=t}^{t+N-1} q({x}_k, {u}_k) \label{mpc:mpcproblem}\\
&{x_{k+1}} = f({x_k, u_k}), \; \forall k \in \{t,..., t+N-1\} \tag{2.1} \\
&{x_0} = {x}_{s} \tag{2.2}\\
&{x_k} \in {X}, \; \forall k \in \{t,..., t+N-1\}  \tag{2.3} \\
\label{mpc:safeste}
&{u_k} \in {U}, \; \forall k \in \{t,..., t+N-1\}\tag{2.4}
\end{align}

The cost function $J$ is made up of a stage cost function $q$ and a terminal cost function $p$ which determine the cost of being at the interim state ${x}_k$ after applying an input ${u}_k$, and the cost of being at the final state ${x}_{t+N}$. The constraints (2.1 - 2.4) assert that the optimisation problem, given by equation~(\ref{model:mpcsystem}), begins from an initial state ${x_s}$ and that the interim state and control inputs  must respect the constraint sets ${X}$ and  ${U}$.

If the dynamics and constraints can be formulated as linear expressions, then the MPC problem can be solved efficiently using standard convex optimisation techniques. However, if the dynamics or constraints are nonlinear, then the problem becomes a nonlinear optimisation problem that is much more computationally challenging to solve. On the other hand, allowing for nonlinear dynamics and constraints may permit one to track complex systems with a higher level of fidelity than using linear expressions. Thus, the computational cost must be evaluated against overall system performance \cite{Seki2002}. 

\subsection{Reachability Analysis}

Reachability analysis is a technique for computing the set of all reachable states of a dynamical system from a set of initial states. The reachable set of ${R}_{t+1}$ can be defined formally as:

\begin{equation}
  {R}_{t+1}(X_0) = f({X}_0, {U})
\label{eq:mpcreachability}
\end{equation}

\noindent where ${X}_0 \subseteq \mathbb{R}^n$ represents the set of initial states, ${U} \subseteq \mathbb{R}^m$ represents the input set. More generally, reachability analysis methods aim to construct a \textit{conservative} flowpipe (\ref{eq:mpcreachtube}) which encompasses all the possible reachable sets of a dynamical system over a time-horizon $[0, T]$. {This can be formalised as follows (in practice the union is computed over a discretised interval):}

\begin{equation}
{R}_{[0, T]}({X}_0) = \bigcup_{t \in [0, T]} {R}_{t}({X}_0)
\label{eq:mpcreachtube}
\end{equation}

Reachability analysis has been widely used in applications that range from the formal verification of systems to problems relating to the safe synthesis of complex systems \cite{doi:10.1146/annurev-control-071420-081941}. The majority of reachability analysis approaches leverage a combination of numerical analysis techniques, graph algorithms, and computational geometry \cite{Asarin2003,Asarin2007}, and while in some cases it is possible to derive the exact reachable set of states, for many classes of systems computing the exact reachable set is infeasible. Thus, deriving the reachable set for these classes of systems involves obtaining a sound approximation (i.e., guarantee to contain a complete reachable set) of this set using a variety of set representations. Consequently, there is an inherent trade-off between the accuracy of the approximation and the time it takes to construct this set. We refer interested readers to the following papers \cite{Asarin2003,Asarin2007} for an in-depth discussion of these techniques.  

\section{Problem Statement and Space Convexification}
\label{sec:mpcproblem}

\begin{figure}[tbp]
  \centering
  \includegraphics[width=0.8\linewidth]{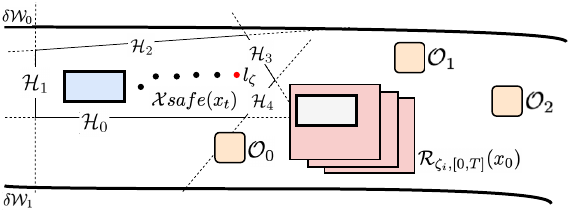}
  \caption{Visualisation of the autonomous racing problem with track boundaries, $\{\delta{W}_0,\delta{W}_1\}$, a dynamic opponent described by its reachable set ${R}_{\zeta_i, [0, T]}(x_0)$ and static obstacles $\{{O}_0,{O}_1,{O}_2\}$. In this figure, the blue rectangle corresponds to the ego vehicle, and the white rectangle corresponds to a dynamic opponent. The main sub-problem is computing an n-number of separating hyperplanes (${H}_0...{H}_4$) which jointly create a polyhedron ${X}_\mathit{safe}$. The computed ${X}_\mathit{safe}$ must contain an ego vehicle and its target location $l_{\zeta}$ as well as not overlap with observable obstacles.}
  \label{fig:mpcreachset}
\end{figure}%

\subsection{Problem Formulation}

In this paper, we consider the general autonomous racing problem \eqref{mpcproblem} where a model predictive controller \eqref{mpc:mpcproblem} is tasked  with generating a sequence of control inputs  $u_{0 \dots T}$ that control a vehicle \eqref{model:mpcsystem} such that it reaches the terminal state ${x}_{T} \in {X}_{f}$ starting from an initial state ${x}_s$ and steering through safe states, where ${X}_{safe}$ and ${X}_{f}$ are the safe states and terminal sets respectively. The goal is to steer the vehicle into the terminal set with the shortest time horizon $T$.

\begin{align}
\begin{split}
& \min_{{T, u_0, u_1,..u_{T-1}}} p({x}_{T}) \: + \: \sum_{k=0}^{T} q({x}_k, {u}_k) \quad s.t. \\
& x_{t+1} = f(x_t, u_t), \quad x_0 = x_s \\
& x_t \in {X}_{safe}, \quad u_t \in {U}, \quad x_{T} \in {X}_{F}
\end{split}
\label{mpcproblem}
\end{align}

In our formulation, the autonomous vehicle operates within a two-dimensional environment ${W} \subset \mathbb{R}^{2}$ enclosed by boundaries $\{\delta{W}_0,\delta{W}_1,\ldots\delta{W}_i\}$ as $\delta{W} \subset {W}$, among a set of dynamic agents $ \zeta = \{\zeta_0,\ldots,\zeta_{i}\}$ ($\zeta$ could be either ego $\zeta_{e}$ or opponent vehicle $\zeta_{o}$) and static obstacles $\{{O}_0,{O}_1,\ldots,{O}_i\}$ with ${O} \subset {W}$. The region of space occupied by a dynamic agent $\zeta_i(x_t) \in W$ in the environment over a time interval $[t, t']$ from its current state $x_t$ is given by its reachable set ${R}_{\zeta_i, [t, t']}(x_t) \subset {W}$. Our assumption is that the static and dynamic obstacles are contained within the two-dimensional environment. Furthermore, we refer to opponent vehicles within the racing environment as dynamic agents and refer to all other dynamic entities as dynamic obstacles.

To obtain a globally optimal solution to problem \eqref{mpcproblem}, as opposed to a locally optimal solution, the model-predictive control problem requires the state-space ${X}$ to be convex. However, because of environment borders, static obstacles and dynamic agents, ${X}$ is generally a non-convex entity. Therefore, the main sub-problem we are addressing in this paper is the computation of the safe, convex and \textit{optimal} state-space ${X}_\mathit{safe}$ in which a safe trajectory starting from $x_0$ to a target location $l_{\zeta} \in {X}_\mathit{safe}$ could be generated using an optimisation-based controller for the autonomous system (see Figure \ref{fig:mpcreachset}). The safe region of the state-space ${X}_\mathit{safe}$ can be defined as follows: 

\begin{equation}
{X}_\mathit{safe} = \{ x \: | \: x \not\in (\delta{W} \cup {O} \cup \bigcup_{i=0}^{N - 1}{R}_{\zeta_i, [0, T]}(x_{0}))\}
\label{mpcsubproblem}
\end{equation}

where $N$ is the number of observable dynamic agents. The computation of ${X}_{safe}$ requires considering only \textit{observable} obstacles, agents and borders. To define observable points we first introduce a notion of the LiDAR sensor which is mounted on the autonomous system and makes it possible to determine the distance to obstacles. The sensor sends $M$ light pulses in an anti-clockwise direction around the autonomous system defined by $\delta \theta$ increments and returns a set of observational points $\{r_0(x_t),..., r_M(x_t)\}$  where a LiDAR observational point $r_i(x_t) \in \mathbb{R}$ in the direction $\theta_i$ can be formally defined in the following way \eqref{eq:mpclidar}:

\begin{align}
\begin{split}
r_i(x_t) = \min_{{O}_i \in {O}}  \min_{z \in {O}_i}  ||z - \zeta(x_t)||_2 \quad 
s.t. \quad \; \atantwo (z - \zeta(x_t)) = \theta_i
\end{split}
\label{eq:mpclidar}
\end{align}

Ranges of the observable LiDAR signals $r_i(x_t)$ can be converted into a two-dimensional point cloud of the ${W}$ where a single point $p_i(x_t)$ of an agent $\zeta(x_t)$ can be defined as a tuple (\ref{eq:mpctuple}):

\begin{equation}
p_i(x_t) = (\zeta(x_t) + r_i(x_t) \cos{\theta_i}, \: \zeta(x_t) + r_i(x_t) \sin{\theta_i})
\label{eq:mpctuple}
\end{equation}



Now, we can define observable static obstacles of $\zeta(x_t)$ as a set ${Q}_{ob}$ of LiDAR points within a constant radius distance $d$ from the agent's state $\zeta(x_t)$:


\begin{align}
\begin{split}
{Q}_{ob} = \{q \; | \; q \in  \{p_0,...,p_{M-1}\}   \land  ||q - \zeta(x_t)||_2 \leq d \}\\
 d \in \mathbb{R}, \; 0 < d \leq \max(r_0(x_t),...,r_{M-1}(x_t))   
\end{split}
\end{align}

Furthermore, the observable unsafe space ${Q}_{ob}$ should include reachable sets of other dynamic agents $\{\zeta_0, \ldots, \zeta_i\}$. However, we are only interested in other dynamic agents which are within some distance $d \in \mathbb{R}^{+}$ and so we update our definition ${Q}_{ob}$ to include reachable regions of other \textit{close} dynamic agents \eqref{eqn:obswreach}: 

\begin{align}
\begin{split}
{Q}_{ob}^{+} = {Q}_{ob} \cup \{q \; | \; q \in \bigcup_{i=0}^{N-1}  {R}_{\zeta_i, [t, t']}(x_t)   
\; \land  \; ||q - \zeta_{e}(x_t)||_2 \leq d\}  
\label{eqn:obswreach}
\end{split}
\end{align}


\subsection{Space Convexification via Separating Hyperplanes}

This paper proposes a solution for the computation of ${X}_\mathit{safe}$ which is based on the convexification of non-convex state space via separating coupled hyperplanes. A hyperplane ${H} = \{x \; | \;  a^{\top}x = b\}$, where $a \in \mathbb{R}^{n}, b \in \mathbb{R}, a \not= 0$,  is a set which splits set $\mathbb{R}^{n}$ into two halfspaces. Let us also denote ${H}^{*}$ \eqref{mpchalfspaces} as one of the halfspaces of the hyperplane ${H}$. A separating hyperplane ${H}$ is then said to separate two disjoint convex sets $A, B$ such that $A \subseteq {H}^{+}$ and $B \subseteq {H}^{-}$ \cite{boydvandenberghe2004}.  
\begin{align}
\begin{split}
&{H}^{*} \in \{{H}^{+}, {H}^{-}\} \quad {H}^{+} \cap {H}^{-} = {H} \\
&{H}^{+} = \{x \; | \;  a^{\top}x \geq b\}    \quad {H}^{-} = \{x \; | \;  a^{\top}x \leq b\}
\label{mpchalfspaces}
\end{split}
\end{align}

An intersection of finite halfspaces is a polyhedron ${P}$ (\ref{mpcpolyhedra}): 
\begin{align}
 {P}_{{H}} = \{x \; | \; x \in \; \bigcap_{i=0}^{N-1}  {H}^{*}_{i} \}
 \label{mpcpolyhedra}
\end{align}

The idea behind a space convexification via separating coupled hyperplanes is to compute a set of hyperplanes ${HS} = \{{H}_0,..., {H}_n \}$ such that together they \textit{create} a polyhedron ${P}_{{HS}}$ which (1) does not intersect with the set observable obstacles of the ego vehicle and (2) the ego vehicle $\zeta_{e}(x_t)$ with its target location $l_{\zeta}$ are within the polyhedron at the time $t$ \eqref{eq:xsafepol}:

\begin{align}
\begin{split}
 {X}_\mathit{safe} = \{ x \: | \:  x  \: \in  \: {P}_{{HS}}  \; \land \: {P}_{{HS}} \: \cap \: {Q}_{ob}^{+} = \emptyset  
  \: \land \: \zeta_{e}(x_t) \in {P}_{{HS}}\  \: \land \:   l_{\zeta(x_t)} \in {P}_{{HS}} \}   
 \label{eq:xsafepol}
\end{split}
\end{align}

The problem of generating a set of separating coupled-hyperplanes ${HS}$ can be defined as an optimisation or satisfiability problem (\ref{mpcoptimisationproblem}) in which $n$ number of hyperplanes are computed such that: 1) each hyperplane separates a part of observable obstacles from the ego vehicle and its target location and 2) all observable obstacles are separated by separating coupled hyperplanes

\begin{align}
\begin{split}
 compute  \quad & {HS} = \{{H}_0,..., {H}_i \}   \quad s.t. \label{mpcoptimisationproblem}  \\
& \forall q^{ob+}_{i} \in {Q}_{ob}^{+} \Rightarrow \exists {H}_i \in  {HS} \; \land \; q^{ob+}_{i} \in {H}^{*}_{i} \; \land \; \zeta(x_t), l_{\zeta(x_t)} \in \mathbb{R}^{n} \setminus {H}^{*}_{i}
\end{split}
\end{align}

The convex and safe polyhedron ${X}_\mathit{safe}$ is the intersection of halfspaces ${H}^{*}_i$ of each hyperplane ${H}_i \in {HS}$ for which $\zeta_e(x_t) \in {H}^{*}_i$ and $l_{\zeta(x_t)} \in {H}^{*}_i$ hold. The problem (\ref{mpcoptimisationproblem}) can be expressed as an optimisation problem on the set of hyperplanes ${HS}$ or polyhedron ${P}_{{HS}}$. One possible \textit{performance} metric could be finding the largest ${P}_{{HS}}$ \cite{10.1007/978-3-642-03367-420,10.1007/BF02187692}.


\section{Autonomous Vehicle Control System}
\label{sec:mpcmethod}
\subsection{Overview of the Closed-Loop Control System}

\begin{figure}[h!]
    \centering
\scalebox{0.85}{
\begin{tikzpicture}
\draw[draw=black, fill=gray!5] (1.7, 0) rectangle (4.5, -4);

\draw[draw=black, fill=gray!5] (5, 0) rectangle (7.5, -.5) node[above, xshift=-13mm] {(Local) Planner};

\draw[draw=black, fill=gray!5] (5, -1) rectangle (6, -3) node[above, xshift=-5mm, yshift=7mm] {${X}_\mathit{safe}$};

\draw[draw=black, fill=gray!5] (6.5, -1) rectangle (7.5, -3) node[above, xshift=-5mm, yshift=7mm] {MPC};

\draw[draw=black, fill=gray!5] (8, 0) rectangle (8.5, -4) node[above, rotate=90, xshift=18mm] {Vehicle Plant};

\node (temp1) at (9, -2) {}; 
\node[below = 25mm of temp1] (temp2) {}; 
\node[left = 85mm of temp2] (temp3) {}; 
\node[above = 15mm of temp3] (temp4) {}; 
\draw[thick] (8.5, -2) -| (temp2);
\draw[thick] (temp2.north) -| (temp4);
\draw[->, thick](temp4.south) -- (1.7, -3.1)  node[xshift=-12mm, above] {Vehicle States};

\draw[draw=black, fill=gray!5] (5, -3.5) rectangle (7.5, -4) node[xshift=-12mm, yshift=2mm] {Reachability};

\node[draw=black, xshift=30mm, yshift=-10mm, minimum width = 24mm, fill=white] (lidar) {LiDAR};
\node[draw=black, minimum width = 24mm, fill=white, below = 4mm of lidar] (rdp) {$\approx$LiDAR};
\node[above = 2 mm of lidar, xshift = -3mm, minimum width = 20mm] (sens) {Sensors};
\node[draw=black, below = 4 mm of rdp, minimum width = 24mm, fill=white] (odom) {Odometry};
\node[draw=black, below = 2.5 mm of odom, minimum width = 23mm, fill=white] (state) {State Estimator};

\node[] (n0) at (0, -1.5) {};
\draw[->, thick] (lidar.south) -- (rdp.north);
\draw[->, thick] (n0) -- (1.7, -1.5) node[above, xshift=-13mm] {Environment};
\draw[->, thick] (6, -2) -- (6.5, -2);
\draw[->, thick]  (5.5, -0.5) -- (5.5, -1);
\draw[->, thick]  (7, -0.5) -- (7, -1);
\draw[->, thick]  (5.5, -3.5) -- (5.5, -3);
\draw[<-, thick]  (5, -.25) -- (4.5, -0.25);
\draw[<-, thick]  (5, -2) -- (4.5, -2);
\draw[->, thick]  (7.5, -2) -- (8, -2);
\draw[<-, thick]  (5, -3.75) -- (4.5, -3.75);
\draw[->, thick]  (4.5, -3.3) -| (7, -3);
\end{tikzpicture}
}
\caption{The architecture of the closed-loop control system for obstacle avoidance}
\label{fig:mpccontrolsystem}
\end{figure}
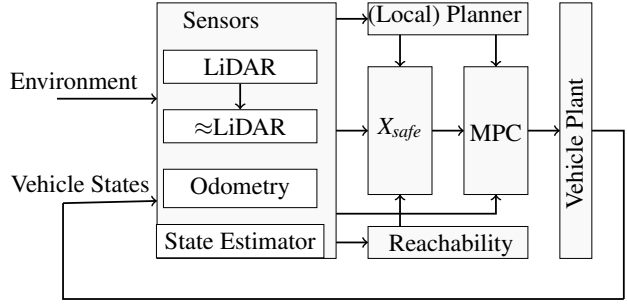

The closed-loop control system for obstacle avoidance which we propose in this paper combines online reachability analysis and non-linear model-predictive control (visualised in Fig.~\ref{fig:mpccontrolsystem}). The control cycle can be divided into four main procedures: sensing, environment data processing and local planning, state-space convexification and solving an optimal control problem. 

The control system relies on the LiDAR sensor to obtain and  identify the set of observable obstacles and safe regions. We then leverage the Ramer-Douglas-Peucker algorithm \cite{doi:10.3138/FM57-6770-U75U-7727} to simplify the observed LiDAR data and reduce the noisiness of its measurements. Doing so allows us to reduce the computation time needed to produce a set of coupled separating hyperplanes. The other sensors, namely, odometry measurements and the results of state-estimators, are used to determine the state of the ego vehicle and other agents respectively. In this work, we assume that the state of the ego vehicle and opponent agents are estimated perfectly. Therefore, we use the ground truth data provided by the simulator. This data is then passed to a (local) planner (e.g., Follow-the-Gap \cite{SEZER20121123}) to select a target position. We then use reachability analysis to compute the set of reachable states for all agents within the environment.

The computation of separating coupled hyperplanes, which produces a safe and convex ${X}_\mathit{safe}$, involves using sensor information, the target location obtained from the local planner, and the set of reachable states of the dynamic agents within the environment. The hyperplanes are then passed to the model-predictive controller, together with the target location and odometry data, which solves an online optimal control problem \eqref{mpcproblem} to determine the optimal inputs for the vehicle.


\subsection{Computing Separating Coupled-Hyperplanes}
\label{sec:mpchyperalgorithm}

The problem of computing separating coupled hyperplanes, which establishes a convex and safe ${X}_\mathit{safe}$, can be formulated as an optimisation (or satisfiability) problem. Thus, we present an optimisation-based method for solving (\ref{mpcoptimisationproblem}) in order to separate the observable obstacle set ${Q}_{ob}^{+}$ from the autonomous system $\zeta_e(x_t)$ and its target location $l_{\zeta_e(x_t)}$ at the state $x_t$. 

In Algorithm \ref{mpchypalgo}, we describe the computation of our separating coupled-hyperplanes ${H}_{0..n}$. First, the set of unsafe states is included in the set ${Q}_{ob}$. The set of unsafe states consists of the set of observable obstacles from the LiDAR sensor and the reachable states of the dynamic agents. Using this set, we then make use of the state of the ego vehicle, the target location obtained from the local planner, and in the case case of the constrained optimisation method a predefined number of hyper-planes to formulate an optimisation problem. Furthermore, only obstacles  (Algorithm  \ref{mpchypalgo} ln. 6) and reachable sets (Algorithm  \ref{mpchypalgo} ln. 7-8) within a distance $d$ from the ego vehicle $\zeta_e(x_t)$ are considered in the hyperplane computation.

\begin{algorithm}\captionsetup{labelfont={sc,bf}}
\caption{The overall algorithm for the computation of separating coupled hyperplanes}
\begin{algorithmic}[1]
\item \textbf{Inputs:} observable radius distance $d \in \mathbb{R}^{+}$ 
\item \textbf{Inputs:} states of the ego vehicle $\zeta_e(x_t) = \{x_e, y_e\}$ and other dynamic agents $\{\zeta_0,\ldots, \zeta_N\}$
\item \textbf{Inputs:} static obstacle data $P = \{p_0,..., p_{N-1}\}$ from the LiDAR (ranges Equation \eqref{eq:mpclidar})
\item \quad Compute target states of the ego vehicle with the local planner $l(x_t) = \{x_t, y_t\}$
\item \quad Compute reachable states ${R} = \bigcup_{i=0}^{N}{R}_{i}$  of observable dynamic agents $\{\zeta_0,\ldots, \zeta_N\}$ 
\item \quad Compute ${Q}_{ob}(x_t)$ by using static obstacle LiDAR data $\{q \: | \: q \in P \land \Vert q - \zeta_e(x_t)\Vert_2 \leq d \}$
\item \quad Compute ${Q}^{+}_{ob}(x_t)$ by combining static and dynamic obstacles ${Q}_{ob}(x_t)\cup {R}$
\item \quad {Encode $q \in {Q}^{+}_{ob}$, $x \in \zeta_e(x_t)$, $x \in l(x_t)$  as constraints of the optimisation problem and solve by using the constrained or bi-level optimisation method}
\item \textbf{ Output: $\{{H}_0,\ldots,{H}_n\}$}
\end{algorithmic}
\label{mpchypalgo}
\end{algorithm}


\noindent \textbf{Constrained Optimisation Method} The first approach uses a derivative-free constrained optimisation formulation which utilises a linear approximation of the objective function and optimisation constraints to solve the aforementioned optimisation problem \cite{Powell2007AVO}. In the optimisation problem, an individual separating hyperplane ${H}_n \in \{{H}_0,..,{H}_{n}\}$ is only \textit{responsible} for separating a subset of ${Q^{+}}_{ob}$ from $\zeta_e(x_t)$ and $l(x_t)$, while the set of all hyperplanes considered should separate  the vehicle from ${Q^{+}}_{ob}$  as a whole. 

For each $q_{ob} \in {Q}^{+}_{ob}$ a separate constraint in the optimisation problem is defined which checks if $q_{ob}$ is separated from the target location and autonomous system $\zeta_e(x_t)$ with some hyperplane ${H}_n$. The constrained optimisation method can use different objective functions which characterise how the set of hyperplanes is derived. For example, the optimisation problem could try minimising the distance between each ${H}_n$ and its associated set of $q_{ob}$, or simply be expressed as a satisfiability problem with a constant objective function. We present an analysis of different optimisation objective functions for this purpose in the evaluation section. \\

\noindent \textbf{Bi-level Optimisation Method}  The problem defined in \eqref{mpcoptimisationproblem} can also be encoded as a bi-level optimisation problem in \eqref{bilevel}. The problem is similar to one solved by Deits and Tedrake \cite{Deits2015} except we are interested in computing a polygon defined by a minimum number of hyperplanes which contains the largest ellipsoids  in the direction of the target location. The \cite{Deits2015} maximises ellipsoid in any possible direction, which is not suitable for the racing context, as the most optimal trajectories produced by the MPC will most likely be along the ego vehicle to the target corridor.

{The outer part of the problem computes the minimum set of separating hyperplanes (the size of the $A$ matrix's diagonal) that separate obstacle points ${Q}^{+}_{ob}$ from the ego vehicle and its target location. The inner part of the bi-level optimisation solves the Chebyshev centre \cite{boydvandenberghe2004} problem\footnote{The reason for maximising the largest inscribable ellipsoid in contrast to directly maximising the area of the safe polyhedron ${X}_\mathit{safe}$ is efficiency. There are no efficient methods for computing the area of irregular polyhedrons, while the Chebyshev centre problem can be solved sufficiently fast.} by finding the centre $q$ of the largest inscribable ellipsoid with radius $R$.}

\begin{align}
\label{bilevel}
\begin{split}
\arg & \min_{A, b}  ||\mathbf{diag}(A)||  \; \text{s.t.} \\
&  A q \geq b, \; \forall q \in {Q}^{+}_{ob} \\
&  A x \leq b, \; \forall x \in\zeta_e(x_t) \cup l(x_t) \\
& \hspace{7mm}\arg  \max_{q, R}   \; R \; \text{s.t.} \\
& \hspace{14mm} a_j q + ||A||R \leq b_j \\
& \hspace{14mm} R \geq 0
\end{split}
\end{align}




\subsection{Reachability Analysis of Dynamic Obstacles}

To perform reachability analysis, we first identify a dynamical model of the vehicle and assume models for the dynamic obstacles within its environment. 

\subsubsection{Dynamic Obstacle Model}

The obstacle-tracking problem is a well-studied and challenging topic within the autonomous vehicle, computer vision, and robotics literature \cite{YilmazObjectTracking}. Typically, some assumptions are required in order to constrain the tracking problem to suit the context of the application. In our framework, we assume that the dynamic obstacles are described by a two-dimensional kinematic model and a corresponding bounding box. The equations describing the kinematic model are given as follows: 
\begin{align*}%
    \Dot{x} & = v_x,%
    \Dot{y}  = v_y%
\end{align*}%
\noindent where $v_x$ and $v_y$ are the velocities in the x and y direction, respectively. Additionally, we make the assumption that we have access to the position and velocity of the other race participants.

While it is possible to use more sophisticated models to describe the behaviour of the dynamic obstacles within the vehicle's environment, for simplicity we selected a two-dimensional kinematic model. However, it is worth noting that there has been a growth in approaches that perform online parameter estimation for dynamic obstacles within a robot's environment through online system identification \cite{Garimella2017}.

\subsubsection{Online Reachability Computation}

 Using the dynamics models obtained in the previous sections, the crux of the real-time reachability algorithm is computing the set of reachable states ${R}_{[0, T]}({X}_0)$ over a finite time horizon. The algorithm utilised within this work is based on mixed face-lifting, which is part of a class of methods that deal with \textit{flow-pipe construction} or \textit{reachtube computation} \cite{Johnson2016}. This is done using snapshots of the set of reachable states that are enumerated at successive points in time, as outlined in Equation~\eqref{eq:mpcreachability}. 

In general, it is not possible to obtain the exact reachable set ${R}_{[0, T]}({X}_0)$, so we compute an over-approximation such that the actual system behaviour is contained within the over-approximation \cite{Lin2020}. The algorithm utilised in this work utilises $n$-dimensional hyper-rectangles (``boxes'') as the set representation to generate reachtubes \cite{Johnson2016}. Over long reach-times, the over-approximation error resulting from the use of this representation can be problematic. However, for short reach-times it is ideal in terms of its simplicity and speed \cite{Bak2014}.

Traditionally, reachability approaches have been executed offline because they are computationally intensive endeavours. However, in \cite{Bak2014,Johnson2016}, Bak et al. and Johnson et al. presented a reachability algorithm, based on the seminal mixed face-lifting algorithm \cite{dang2000}, capable of running in real-time on embedded processors. The algorithm is implemented as a standalone C-package that does not rely on sophisticated (non-portable) libraries, recursion, or dynamic data structures and is amenable to the anytime computation model in the real-time scheduling literature. In this regime, each task produces a partial result that is improved upon as more computation time is available, known as an anytime algorithm \cite{Johnson2016}. We refer readers to the following papers for an in-depth treatment of these procedures \cite{dang2000,Bak2014,Johnson2016}.

\section{Evaluation}
\label{sec:mpccasestudy}

In this section, we present a runtime analysis of proposed algorithms for computing separating coupled hyperplanes and an evaluation of the overall control system by using the F1TENTH simulation platform. In the following section, we first describe an optimisation-free method (MPCC) for computing separating hyperplanes, which will be used to compare against our proposed approaches.

\subsection{MPCC Optimisation-free Hyperplane Approach} {In \cite{Liniger2014} Liniger et al. tackled the autonomous racing problem via a nonlinear MPC problem that encoded the obstacle avoidance problem by means of a high-level corridor planner based on dynamic programming. The safe corridor that their framework utilised was constructed by projecting the points along the centre line of the track onto the racetrack borders (one for the left border, and one for the right border). Their regime demonstrated success in controlling 1/43 scale race cars, driven at speeds of more than 3 m/s using controllers executing at 50 Hz sampling rate on embedded computing platforms \cite{Liniger2014}. While their evaluation was limited to environments with static obstacles, we experimented with using such a scheme to obtain the separating hyperplanes framing our MPC problem.  We refer readers to the following paper for an in-depth discussion of their approach \cite{Liniger2014}.}

\begin{figure}
    \begin{minipage}{0.49\textwidth}
        \includegraphics[width=1\textwidth]{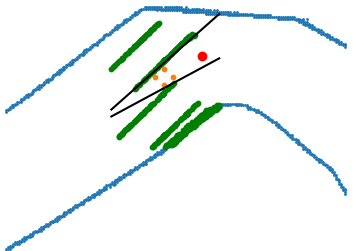} 
        \subcaption{{Constrained Optimisation Method}}
    \end{minipage}\hfill
    \begin{minipage}{0.49\textwidth}
        \includegraphics[width=1\textwidth]{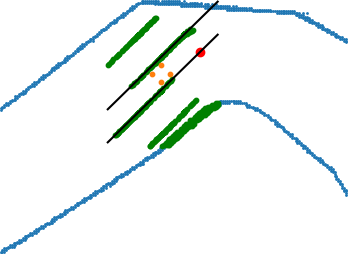} 
        \subcaption{MPCC approach}
    \end{minipage}
            \caption{A snapshot of the artificial overtaking scenario with two opponents represented: combined observable static and dynamic obstacles ${Q}_{ob}^{+}$ (green points), corners of the ego vehicle (orange points), target point (red point), computed hyperplanes (black) and the boundaries of the racetrack (blue).}
            \label{eval:snapshot}
\end{figure}

\subsection{Offline Analysis of Convexification Algorithms}
\label{sec:mpcruntimecomp}

Deploying optimisation-based methods into real-time autonomous control systems requires careful consideration of timing constraints issued by the optimisation method. The computation time of the proposed methods can be affected by the number of obstacle points being considered or in the case of the constrained optimisation method by the selected optimisation cost function. {In these experiments, we aim to evaluate the \textit{quality} of separating hyperplanes generated by different approaches and the computation time. The former is assessed by inscribing the largest   circle with radius $R$ between generated hyperplanes (the centre of the circle must be between the ego vehicle and its target location) and gives a reasonable size approximation of the $X_{safe}$ in the travelling direction of ego vehicle. In the first set of experiments, we traversed the ego vehicle along a predefined path on one of the two racetracks: Porto (see Figure \ref{fig:mpcmpcc}) and Walker without other dynamic agents, and then considered an artificially created overtaking scenario with one and two opponents (visualised in Figure \ref{eval:snapshot}). The results of this experiment are summarised in Table \ref{tab:mylabel}. The MPCC approach is clearly more time efficient compared to our proposed approaches as it is not an optimisation-based approach. However, it produces a smaller average inscribed circle radius (i.e., smaller $X_{safe}$), particularly, in the overtaking scenarios with up to 16 per cent smaller $R$ in comparison to the largest averaged $R$. The bi-level optimisation approach is around tenfold faster than the constrained optimisation approach, as its outer problem is a linear programming problem, which can be efficiently solved even with a larger number of obstacles (problem constraints). However, our experiments show that the bi-level optimisation method does not always produce the largest $X_{safe}$ and in some cases generates more than two hyperplanes, which would negatively affect solving the MPC problem.

In the second experiment, we increased the number of generated (randomly positioned) obstacle points around an ego vehicle to evaluate our method's scalability with respect to a larger number of obstacles. For the constrained optimisation method we considered three types of objective functions: Hausdorff, Euclidean distances and a satisfiability problem which only requires satisfying optimisation constraints. For each objective function and bi-level optimisation, the number of obstacle points varied from 10 to 2000. The evaluation results are visually shown in Figure \ref{eval:mpcresults}.


\begin{table}[h]
\centering
\begin{tabular}{c  c c c  c}
\textbf{Experiment Scenario} \quad & \quad \textbf{Approach} \quad & \quad ${\mathbf{H}}$ \quad & \quad \textbf{Time (s)} \quad & \quad $\mathbf{R}$ \\
\hline
\hline
Porto (w/o obstacles) \quad  & \quad MPCC \quad & \quad 2 \quad  & \quad \textbf{6.46} $\mathbf{\times 10^{-5}}$ \quad & \quad 1.29 \\
Porto (w/o obstacles) \quad  & \quad Constrained Optimisation \quad & \quad 2 \quad  & \quad 0.132  \quad & \quad \textbf{1.30}\\
Porto (w/o obstacles) \quad  & \quad {Bi-level Optimisation} \quad & \quad 2.16 \quad  & \quad 0.019  \quad & \quad 1.072\\
Walker (w/o obstacles) \quad  & \quad MPCC \quad & \quad 2 \quad  & \quad $\mathbf{7.65 \times 10^{-5}}$ \quad & \quad 0.702 \\
Walker (w/o obstacles) \quad  & \quad {Constrained Optimisation } \quad & \quad 2 \quad  & \quad 0.12 \quad & \quad 0.675\\
Walker (w/o obstacles) \quad  & \quad {Bi-level Optimisation} \quad & \quad 2.26 \quad  & \quad 0.019  \quad & \quad \textbf{0.715}\\
\hline
Overtaking (1 opponent) \quad  & \quad {MPCC} \quad & \quad 2 \quad  & \quad $\mathbf{8.679 \times 10^{-5}}$  \quad & \quad 0.627\\
Overtaking (1 opponent) \quad  & \quad {Constrained Optimisation} \quad & \quad 2 \quad  & \quad 0.18  \quad & \quad 0.661\\
Overtaking (1 opponent) \quad  & \quad {Bi-level Optimisation} \quad & \quad 2.13 \quad  & \quad 0.022  \quad & \quad \textbf{0.755}\\
\hline
Overtaking (2 opponents) \quad  & \quad {MPCC} \quad & \quad 2 \quad  & \quad 9.79 $\times$ 10$^{-5}$  \quad & \quad 0.465 \\
Overtaking (2 opponents) \quad  & \quad {Constrained Optimisation} \quad & \quad 2 \quad  & \quad 0.265 \quad & \quad \textbf{0.488}\\
Overtaking (2 opponents) \quad  & \quad {Bi-level Optimisation} \quad & \quad 2.59 \quad  & \quad 0.026 \quad & \quad 0.473\\

\end{tabular}
\caption{Offline evaluation of different methods for computing separating hyperplanes with average computation time, an average inscribed radius $R$ and an average number of hyperplanes $H$. }
\label{tab:mylabel} 
\end{table}

\begin{figure}[!h]%
    \centering
\includegraphics[width=0.65\paperwidth]{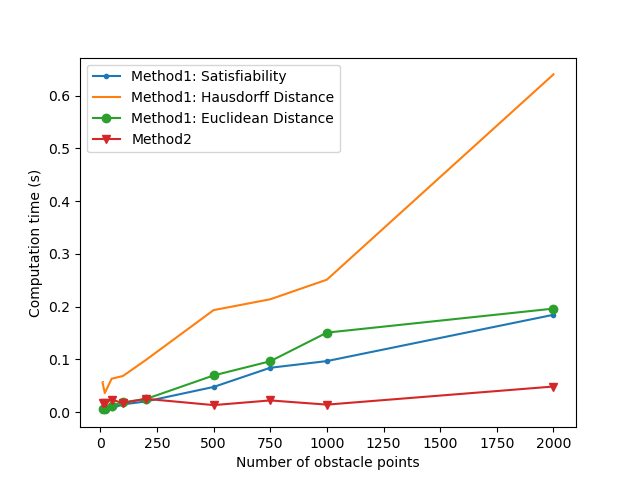}
    \caption{Offline evaluation of separating coupled hyperplane computation time against different numbers of obstacle points (optimisation constraints), different objective functions (method 1 - constrained optimisation approach, method 2 - bi-level optimisation).  }
    \label{eval:mpcresults}
\end{figure}%


\subsection{Real-Time Control System Evaluation} 

Our real-time evaluation of the overall control system (MPC Hype) includes a diverse set of experiments that include changing the number of racing agents present within the racetrack, including additional dynamic obstacles within the racetrack, adding static obstacles onto the racetrack, and changing the racing environment. We compare the performance of our approach against a set of controllers typically utilized within the F1TENTH racing competitions with respect to two metrics that we refer to as \textit{efficiency}, and \textit{safety}. \textit{Efficiency} is the total distance that the F1TENTH vehicle traverses around the track divided by the amount of time it took to do so.\footnote{This is equivalent to the average speed attained during the experiment.} \textit{Safety} corresponds to the controller's ability to avoid collisions over a set of experimental runs (i.e., 10 collisions in 20 experiments corresponds to a safety score of 50\%). The following controllers were utilised as a local planning mechanism for selecting the target point used in our MPC regime. Additionally, we utilised them as a baseline comparison for our approach. 

\paragraph{Pure Pursuit}

The Pure Pursuit algorithm is a widely used path-tracking algorithm that was originally designed to calculate the arc needed to get a robot back onto a path \cite{Coulter-1992-13338}. It has shown great success in being used in numerous contexts, and in this work, we utilise it to design a controller that allows the F1TENTH vehicle to follow a path along the centre of the racetrack.

\paragraph{Gap Following} Obstacle avoidance is an essential component of a successful autonomous racing strategy. Gap following approaches have shown great promise in dealing with dynamic and static obstacles. They are based on the construction of a gap array around the vehicle used for calculating the best heading angle needed to move the vehicle into the centre of the maximum gap \cite{okelly2020}. In this work, we utilise a gap following controller called the ``disparity extender'' by Otterness et al. that won the F1TENTH competition in April of 2019 \cite{otterness2019}. 

\begin{figure}[!htbp]
  \begin{center}
    \includegraphics[width=0.99\linewidth]{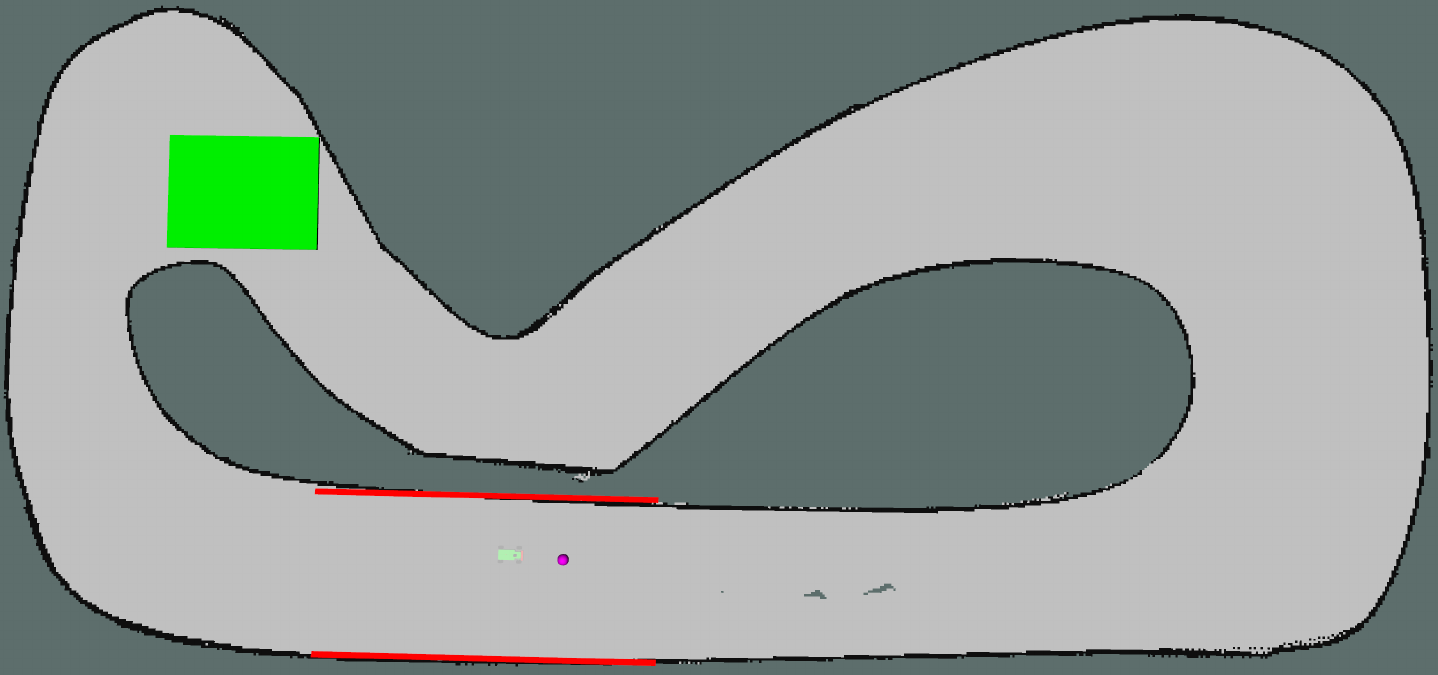}
   \caption{An example of a two-agent racing scenario. The bright green rectangle, represents the reachable set (convex hull) of the opponent vehicle over a $t=0.5$ second time horizon, while the faded green vehicle represents the ego vehicle. The purple dot corresponds to the target location obtained from the local planner. The red lines are the two parallel half spaces that approximate the traversable region within the racetrack.}
  \label{fig:mpcmpcc}
   \end{center}
\end{figure}

\begin{table*}[h!]
    \centering
    \caption{Performance summary of two-car experiments (without static obstacles): DE (Disparity Extender), PP (Pure Pursuit), MPCC (Model-Predictive Control with Contouring) and MPC Hype (our control system with the constrained optimisation hyperplane computation)}
    \label{tab:results}
\resizebox{0.95\linewidth}{!}{
\begin{tabular}{ccccccc}
  &           &                                   &                 &                      &         \\
Track &  Approach  & Local Planner &  Ego Efficiency (m/s) &  Opponent Efficiency (m/s) & Race Duration (s) & Safety (\%) \\
\hline
Porto  &            DE &                              DE &            5.29 &                 4.65 &              51.57 &   38.33 \\
Porto  &      MPC Hype &                              DE &            0.00 &                 5.27 &               5.53 &    0.00 \\
Porto  &      MPC Hype &                              PP &            3.06 &                 5.18 &              25.74 &   13.33 \\
Porto  &          MPCC &                              DE &            3.00 &                 4.97 &               7.12 &   20.00 \\
Porto  &          MPCC &                              PP &            3.00 &                 5.34 &              55.14 &   46.67 \\
Porto  &            PP &                              PP &            4.70 &                 5.33 &              60.0 &  100.00 \\
\hline
Porto  &            DE &                              DE &            5.38 &                 4.10 &              33.78 &   28.33 \\
Porto  &      MPC Hype &                              DE &            1.19 &                 4.50 &               5.40 &    0.00 \\
Porto  &      MPC Hype &                              PP &            2.75 &                 2.96 &              43.26 &   30.00 \\
Porto  &          MPCC &                              DE &            1.66 &                 4.23 &               5.39 &    3.33 \\
Porto  &          MPCC &                              PP &            1.83 &                 4.00 &               5.37 &   16.67 \\
Porto  &            PP &                              PP &            4.70 &                 3.73 &              57.30 &   70.00 \\
\hline        
\end{tabular}}
\end{table*}%

Our evaluation included a sizeable diversity of experiments with respect to the number of vehicles present in the racing environment, the presence of static and dynamic obstacles, the racetrack used for the autonomous race, the local planner chosen to select goal points, and the method selected to obtain the separating hyperplanes. {Each configuration was evaluated over 30 experimental runs of 60 seconds. Table \ref{tab:results} displays the results of experiments with two and three cars respectively (separated by horizontal line) on a single track without static obstacles (a screenshot of the two agent experiment is shown in Figure \ref{fig:mpcmpcc}). In the table that follows, DE corresponds to the disparity extender, PP corresponds to pure pursuit, MPCC corresponds to the approach presented by Liniger et al. \cite{Liniger2014}, and MPC Hype corresponds to the optimisation-based approach presented in this document. Finally, Race Duration corresponds to the amount of time the agents were able to race before a collision occurred. 

The results from the experiments suggest that our proposed control system (MPC Hype) can increase autonomous vehicle safety without loss of efficiency (compared to MPCC), especially when the number of opponent vehicles is increased. However, results also suggest that the performance of our MPC implementation could be further improved, for example, by improving the MPC cost function to generate better speed profiles in corners. This would also provide us with more evidence of the hyperplane approach when ego velocity is increased. Furthermore, our experimentation setup did not differentiate between different types of collisions, for example, collisions, where the opponent vehicle collided with the back of the ego vehicle and the reverse situation, were treated equally (i.e., counted the same in the safety metric). A more nuanced safety metric with a \textit{blame} factor would provide a better understanding of our control system performance. 
\section{Conclusions and Future Work}
\label{sec:mpcconclusions}

This paper presented an optimisation-based approach for static and dynamic obstacle avoidance problems within an autonomous vehicle racing context. Our control regime leveraged online reachability analysis and sensor data to compute the maximal safe traversable region that an agent can traverse within the environment. We described a technique for computing a convex safe region via a novel coupled separating hyperplane algorithm. This derived safe area was then used to formulate a nonlinear model-predictive control problem that sought to find an optimal and safe driving trajectory with varying degrees of efficacy. Our experimental evaluation demonstrated that our approach was feasible as an obstacle avoidance strategy. Finally, we assessed the runtime requirements of our proposed approach by analysing the effects of a set of varying optimisation objectives for generating these coupled hyperplanes. 

There are a number of future work directions we would like to explore. Firstly, our study did not consider uncertainty in sensors, our future work will seek to include uncertainties arising from the state estimation of opponent vehicles in their reachable set computation. Secondly, future studies would include an analysis against hierarchical control architectures that decompose the obstacle avoidance problem into planning and trajectory tracking. Lastly, we would like to evaluate the proposed approach on the physical F1TENTH platform in order to validate further that our approach admits low resource requirements.

\bibliographystyle{eptcs}
\bibliography{generic}
\end{document}